# An Empirical Bayes Approach to Regularization Using Previously Published Models

D.K. Smith; L.E. Smith; B. Kroncke; F.T. Billings; J. Meiler; J.D. Blume

October 10, 2017

Derek Smith, Department of Biostatistics
2525 West End, Ste.11000 Nashville, TN 37203.
E-mail: derek.smith@vanderbilt.edu

# 1 Abstract

This manuscript proposes a novel empirical Bayes technique for regularizing regression coefficients in predictive models. When predictions from a previously published model are available, this empirical Bayes method provides a natural mathematical framework for shrinking coefficients toward the estimates implied by the body of existing research rather than the shrinkage toward zero provided by traditional L1 and L2 penalization schemes. The method is applied to two different prediction problems. The first involves the construction of a model for predicting whether a single nucleotide polymorphism (SNP) of the KCNQ1 gene will result in dysfunction of the corresponding voltage gated ion channel. The second involves the prediction of perioperative serum creatinine change in patient's undergoing cardiac surgery.

Empirical Bayes ; Penalization ; LASSO

# 2 Introduction

Often, we are faced with the challenge of building predictive models that are specific to a given purpose. When there exists a previously published model that was fit for a similar purpose it may contain valuable information to inform our predictions, but if it was not designed for the particular research objectives the investigators wish to pursue or if it was trained in a different population the behavior of the model needs to be verified and possibly updated for the new context. When data from the new context are available this reassessment can be done directly, but it is also possible that by updating the model for performance in our specific situation of interest we could obtain better predictive performance. However, it is not always clear how to extract information from previously published models to inform the training of a new model, especially when the published details on the model are vague. We refer to these models as 'black box models' when either no information is available on how inputs are transformed into predictions or that information is incompatible with the type of model we want to use for the new data.

In this manuscript an empirical Bayes approach toward updating published black box predictive models is presented. This method uses cross-validation to decide how much emphasis should be put on the situation-specific data versus how much should be put on the previously published model. This emphasis tradeoff provides a natural regularization that acts similarly to other penalization schemes.

One place in which this situation has become commonplace is in the prediction of the pathogenicity of a single nucleotide polymorphism (SNP). Most of the published models are trained using mutations in a wide variety of genes. The genes in turn encode a variety of proteins, the dysfunction of which can lead to disease. The resulting dysfunction is defined quite generally to mean any disease process arising from substitution of an amino acid regardless of the nature the resulting disease process. These models are typically evaluated either via cross-validation in the same dataset in which they are fit or in similar holdout datasets composed of many SNPs in many genes.

An investigator who requires a predictive model for forecasting whether variants of one particular gene result in a specific disease process might find that these models do not demonstrate predictive performance commensurate with what they displayed when being evaluated in heterogeneous datasets in this new gene and disease specific context. However, with gene-specific data in hand it may be desirable to create a new gene-specific model which still leverages the information about amino acid substitutions generally. Not only could the investigator have a greater degree of confidence in the model's applicability, but a degree of model interpretability could be recovered.

Another common problem in predictive modeling is covariate selection and managing the model's degrees of freedom. One way of handling situations where information is limited relative to the model being considered is to regularize or "shrink" model coefficients. The general principle behind regularization is to bias the estimated coefficients in order to reduce their variance. This regularization allows for the fitting of more stable models with larger numbers of covariates and often results in better predictive performance.

An example is presented where data on the relationship between SNPs of the KCNQ1 gene and the development of long QT syndrome are examined to determine how they can best be utilized in conjunction with previously constructed models. The information on KCNQ1 is integrated into six previously published models predicting protein dysfunction and the various models are assessed. The predictive performance of these models is then compared to an optimized elastic net.

A second example is presented in which the primary aim is to forecast whether cardiac surgery patients will develop perioperative acute kidney injury (AKI). In this example, the dataset is augmented by using predictions from the NSQIP surgical risk index. The NSQIP model was designed to predict the probability of any surgical complication arising including renal complications along with other unrelated complications. The empirical Bayes regularized model was again benchmarked against the elastic net.

# 3 Empirical Bayes Method

## 3.1 Regularization and Bayesian Methods

Bayesian models have many advantages due to their inherent regularization toward values that are favored by the prior. A well-chosen prior can provide a great benefit to predictive performance, particularly when data is scarce or the number of parameters is large compared to the available data. Similar benefits are enjoyed by penalized regression models such as ridge regression (L2 penalty) [1], LASSO regression (L1 penalty) [2], and elastic net (combination of L1 and L2 penalties) [3] with regard to situations with large numbers of parameters.

Penalized regression involves the optimization of an objective function. For example, ordinary linear regression involves finding
$$argmin_\beta \left\{(y - \mathbf{X}\beta)^2\right\},$$
which involves no penalty. In ridge regression the objective function becomes
$$argmin_\beta \left\{(y - \mathbf{X}\beta)^2 + \lambda \sum \beta^2\right\}.$$

The $\lambda$ parameter is a tuning parameter that is selected by cross-validation. It governs the amount of shrinkage applied to the model coefficients. As the model coefficients are penalized for being large, shrinkage in each of ridge, LASSO, and elastic net regression biases the coefficients toward zero in exchange for reduced variation in estimating the coefficient.

Any gains seen in predictive performance resulting from introducing a penalty are no coincidence. The ridge regression is regularizing in exactly the same way as a Bayesian model with a independent normal priors with mean of zero. The dispersion of the normal prior determines the degree of regularization in the same way the tuning parameter does for the penalized regression. Similar arguments can be made for LASSO being a special case of a Bayesian model with a Laplace prior [4].

Each of these approaches to regularization has benefits and limitations. Bayes methods provide an inherent degree of regularization biasing coefficient estimates towards a coherent value for the parameter that is based on prior observations. However, the degree of regularization is related to the dispersion of the prior and not selected for optimal predictive performance as is commonly done with other penalization methods. Conversely, $L^p$ penalizations and their generalizations like elastic net attempt to optimize the degree of regularization, but they bias the coefficients toward an arbitrary location, namely zero. Although biasing coefficients toward zero may provide some advantages for inference or when sparsity is desirable, biasing them toward coherent values can provide greater gains in predictive performance since not only is the variance reduced by shrinking toward a static location but the bias introduced is in a meaningful direction.

It is usually preferable that Bayesian priors be based on actual prior knowledge but in many cases the pertinent information is not readily accessible. When priors are based on real knowledge, the Bayesian approach can be still be viewed as a regularization procedure in addition to its other benefits. Resulting coefficient estimates will still demonstrate bias, in frequentist terms, and a reduction in variance. However, instead of biasing the estimates toward zero the Bayes procedure shifts them toward the most plausible values in the prior. The empirical Bayes procedure is an attempt to make use of this same commonsense regularization in a situation where information is either not available or its applicability is questionable.

For instance, in the KCNQ1 example the previously published models are based on neural networks, support-vector machines and other algorithms using a variety of different features. Even with detailed information about these models coming up with plausible values for logistic regression coefficients would be difficult. Knowing how diffuse a prior about such a plausible value would need to be to achieve optimal predictive performance adds another layer of complexity ultimately making a proper Bayesian solution untenable.

The empirical Bayes method presented here circumvents these difficulties by estimating the "prior" from the data. This data driven solution often yields a "posterior" that underestimates the uncertainty associated with the coefficient estimates, but these traditional criticisms of empirical Bayes methods are less relevant with respect to their implementation in predictive methods in which prediction uncertainty is not a primary concern. Although these methods share the Bayesian machinery, they are in many cases better thought of as a system for regularization.

## 3.2 Setup

Consider the following situation. One set of researchers is studying a disease process in a particular population $\mathcal{P}$. These researchers collect a sample from their population consisting of $(Z, y)$ where $Z$ is a matrix of predictors and an outcome variable $y$. Fitting their predictive model of choice the researchers estimate a function

$$\hat{y} = h(Z), which\ typically\ takes\ the\ form\ of\ \hat{E}[Y_m|Z].$$

Now suppose there is a second group of researchers who are interested in studying a very similar problem in population $\mathcal{P}'$, but their sample size is limited so regularization is desirable. In order to develop the best predictive model possible these researchers intend to leverage the existing model to the extent that it generalizes to their population $\mathcal{P}'$. The second group of researchers collects data on the features they believe to be relevant to the prediction problem, $\mathbf{X}$, which may or may not differ from those chosen by the first group. However it is assumed that there is some overlap due to the closely related nature of the problem, $\mathbf{X} \cap \mathbf{Z} \neq \emptyset$. For our purposes we presume the researchers intend to use a generalized linear model such that

$$E[Y|X] = g^{-1}(\beta_0 + \mathbf{X}\beta),$$

although this restriction is not necessary generally. What follows is a description of an empirical Bayes technique to synthesize a new model specific to the $\mathcal{P}'$ population.

## 3.3 Empirical Bayes Prior Estimation

Using the previously published model to estimate an empirical Bayes "prior" provides a way to marry the shrinkage toward a coherent value of the Bayes approach with the data-driven solution for the optimal amount of regularization employed by the penalized regression techniques. Presuming the model, $h()$, is a blackbox with different features from the proposed model additional steps are required to estimate plausible values for the $\beta$s, as they can not be simply read from the published manuscript.

One way to arrive at a plausible prior estimate is through the use of a parametric bootstrap based on $h(Z)$. This is especially useful for classification problems where $Y$ is categorical such that a logistic or probit model are natural choices for the researcher. In cases where $h(Z)$ is the same data type as $Y$ plausible values could come from fitting

$$g(h(Z)) = \beta_0 + X\beta.$$

So if $h(Z)$ and $Y$ are both continuous a GLM can be fit directly using $h(Z)$ as the outcome. The coefficient estimates could then be used as centrality parameters in the empirical Bayes prior. In the case of binary data this approach cannot be used because the new model cannot be fit directly, i.e. $h(Z)$ is continuous on $[0, 1]$ and not appropriate as the outcome in a logistic model.

The parametric bootstrap approach, however, is quite general. The amount of useful information extracted from the previously published model will vary based on the quality of the assumed parametric model. The parametric bootstrap procedure begins with the assumption

$$Y|Z \sim F_Y(y|\mu = E[Y|Z]) \approx \hat{F}_Y(y|\mu = h(Z))$$

that the distribution of Y can be well approximated using the information available from the previously published model. For example in the case of binary $Y$, the assumption is

$$Y|Z \sim Bern(p) \approx Bern(h(Z)).$$

Using this estimated distribution for $Y$, new outcomes are generated, $Y^*$. Given that the estimated distribution was appropriately selected, $Y^*$ now has the proper data type to be fit by the researcher's proposed GLM

$$g(Y^*) = \beta_0 + X\beta.$$

This process is repeated several thousand times with $(\beta_0, \beta)$ being recorded for each iteration. These samples can then be used to estimate a prior for the empirical Bayes GLM.

When selecting a family for the prior there are some important considerations. As the prior is primarily a means to regularize the model, it is convenient to be able to alter the scale of the prior separately from the location, so families that have independent location and scale parameters are advantageous. Some examples include independent Gaussian, t, and Cauchy distributions. Recall that with other $L^p$ regularized regressions the resulting shrinkage is toward zero and the amount of shrinkage is regulated by the tuning parameter $\lambda$, Figure 1. In empirical Bayes regularization the shrinkage will occur toward the center of the chosen (presumably unimodal) prior, and the degree of the shrinkage will be governed by the scale parameter. Therefore when assessing the performance of the empirical Bayes model, the scale should be used as a tuning parameter and selected by cross-validation. Figure 2 displays the regularization path for the empirical Bayes model. Note that with the LASSO penalty coefficients 3 and 4 are quickly eliminated whereas under empirical Bayes regularization they slightly increase over the path. There is also a marked difference in the regularization of components 1 and 2.

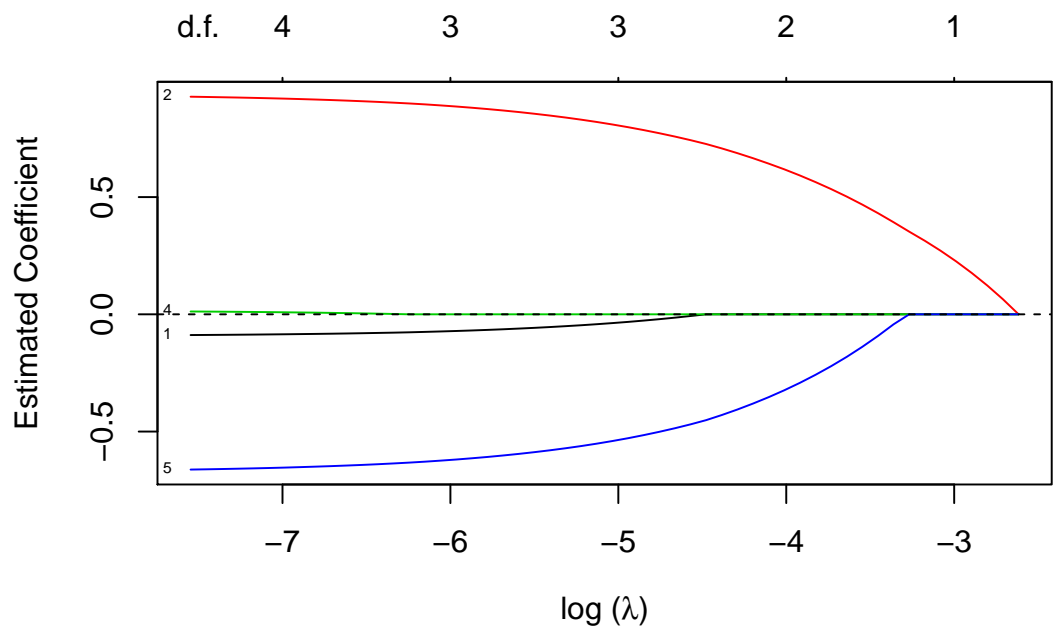

Figure 1: Regularization path for the coefficients of the LASSO penalized logistic regression relating protein dysfunction to the first five principle components of the selected predictors in the KCNQ1 data. Note that component 3 was regularized to zero for the entire displayed path.

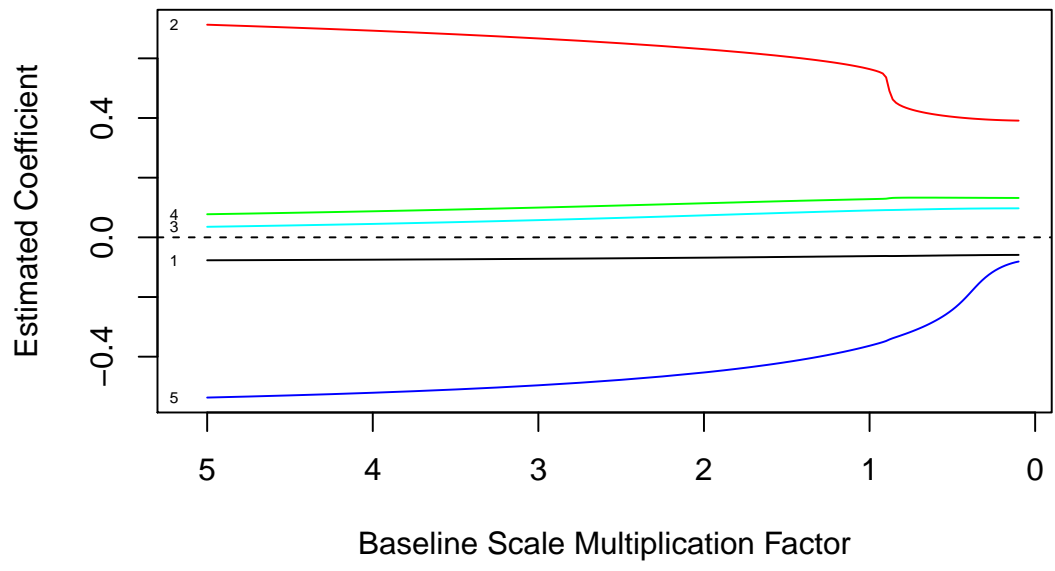

Figure 2: Regularization path for the coefficients of the empirical Bayes logistic regression relating protein dysfunction to the first five principle components of the selected predictors in the KCNQ1 data. This model used a prior provided by SNAP.

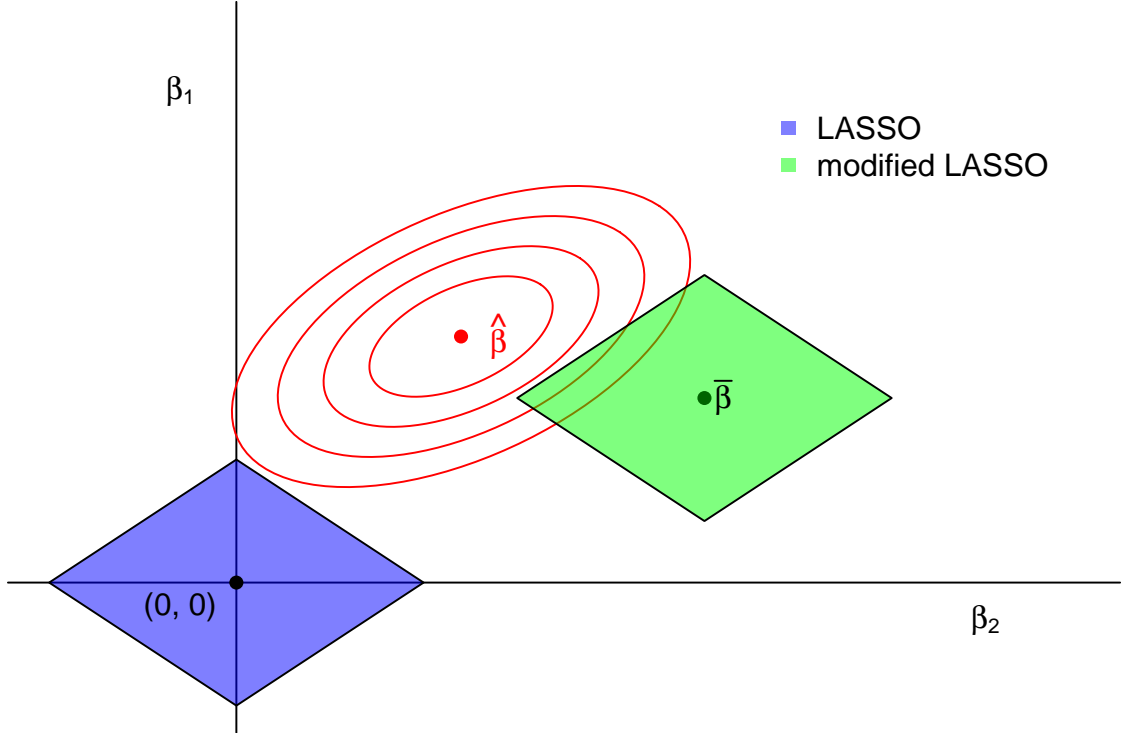

Figure 3: Constraint regions for the LASSO and modified LASSO penalties

It is worth noting that the regression path suggested here is not the only choice that could have been made. Another equally valid way to use the bootstrap estimates of centrality, $\bar{\beta}$, is to incorporate them in an $L^p$ regularization objective function. For example ordinarily the LASSO objective function for a binary logistic model is given by

$$argmin_\beta \left\{ -\frac{1}{n} \sum_{i=1}^{n} \left( y_i(\beta_0 + x_i\beta) - \log(1 + e^{\beta_0 + x_i\beta}) \right) + \lambda \sum_{k=1}^{p} |\beta_k| \right\}.$$

With a slight modification, rather than penalizing the absolute magnitude of the coefficient penalization can occur on the absolute difference between the coefficient value suggested by the published model and the value suggested by the situation specific data. In this case the objective function becomes

$$argmin_\beta \left\{ -\frac{1}{n} \sum_{i=1}^{n} \left( y_i(\beta_0 + x_i\beta) - \log(1 + e^{\beta_0 + x_i\beta}) \right) + \lambda \sum_{k=1}^{p} |\beta_k - \bar{\beta}_k| \right\}.$$

As the unmodified LASSO solution is akin to a Bayes solution incorporating a Laplace prior centered at zero, the modified LASSO solution is akin to an empirical Bayes solution incorporating a Laplace prior where the centrality parameter is estimated based on the previously published predictive model. This modification amounts to translating the traditional LASSO constraint region by a factor of $\bar{\beta}$, see figure 3.

# 4 KCNQ1 Example

## 4.1 Dataset

The available data for KCNQ1 specific SNPs consists of 237 observations which were obtained from either an in-house curated dataset or from ClinVar [5]. Variants having in ClinVar which had conflicting annotations were omitted. Only 17 of the 237 SNPs were non-deleterious limiting the effective sample size.

## 4.2 Published Models

Several previously published models were considered for comparison including: MAPP [6], PHD-SNP [7], Polyphen-2 [8], PredictSNP [9], SIFT [10], and SNAP [11].

## 4.3 Statistical Modeling

In this example the outcome relates to the probability a mutation is associated with long QT syndrome. This is a binary outcome and as such an empirical Bayes logistic model was chosen for this experiment. Features were selected by subject matter experts *a priori*, but the data was far too limited to accommodate them all. Dimensionality reduction and orthogonalization of the predictors was achieved by taking the first five principle components. The identified predictors included evolutionary information, changes due to the amino acid exchange, and structural information.

Independent Cauchy distributions were used as the prior as they have shown some beneficial properties in logistic regression [12] and satisfy the criteria of being a location-scale family. The probability of protein dysfunction was extracted from online prediction tools, and was recalibrated to give appropriate probabilities in the new population. This was achieved by fitting the regression $g(Y) \sim ns(h(Z))$, where $ns()$ represents a natural spline. The recalibrated scores will be denoted, $h_c$.

The parametric bootstrap procedure outlined in the previous section was then used to generate $Y^* \sim Bern(h_c)$. A logistic regression with the same model matrix as that being proposed was then fit and the coefficients were stored. This process was repeated 3,000 times. These empirical prior estimates were then replaced by a Cauchy distribution with its center at the median value. A baseline scale was assigned to each feature based on the empirical prior's dispersion. Due to the limited amount of data, stratified cross-validation took place in a repeated two-fold manner with 150 replicates.

The cross-validated Matthews' Correlation Coefficient (MCC) was used as the primary measure of predictive performance as it is very commonly used in this field's literature. Figure 4 shows the resulting MCC when the scale of the priors is scaled from its baseline value with some of the models such as the MAPP and SNAP favoring the prior distribution and others like SIFT and PHD-SNP favoring the data slightly. In each case predictive performance is improved by employing the empirical Bayes regularization. For comparison, an optimized elastic net model with $\alpha = 0.95$ and $\lambda = 0.017$ put through the same cross-validation procedure produced a cross-validated MCC of only 0.305, substantially less than those demonstrated by empirical Bayes regularization in all but the SIFT model.

The benefits of utilizing empirical Bayes regularization can be further emphasized by considering the case where the previously published model is entirely noise. Random predictions were generated for each mutation in the KCNQ1 dataset.

$$h_{rand}(Z) \sim Unif(0,1)$$

The same empirical Bayes procedure was carried out using the centrality parameters suggested by this purely noise existing model and optimizing via the scale of the empirical Bayes prior, Figure 5. It is clear in this figure that despite $h_{rand}$ contributing no worthwhile information to the problem, the regularization still benefited the predictive performance in the same way other $L^p$ regularizations would have done. However, in this case no additional benefit is achieved as the location to which the coefficients are being shrunk is no longer being chosen in a scientific way. As more weight is put on the coefficient values suggested by $h_{rand}$ we see predictive performance decline dramatically.

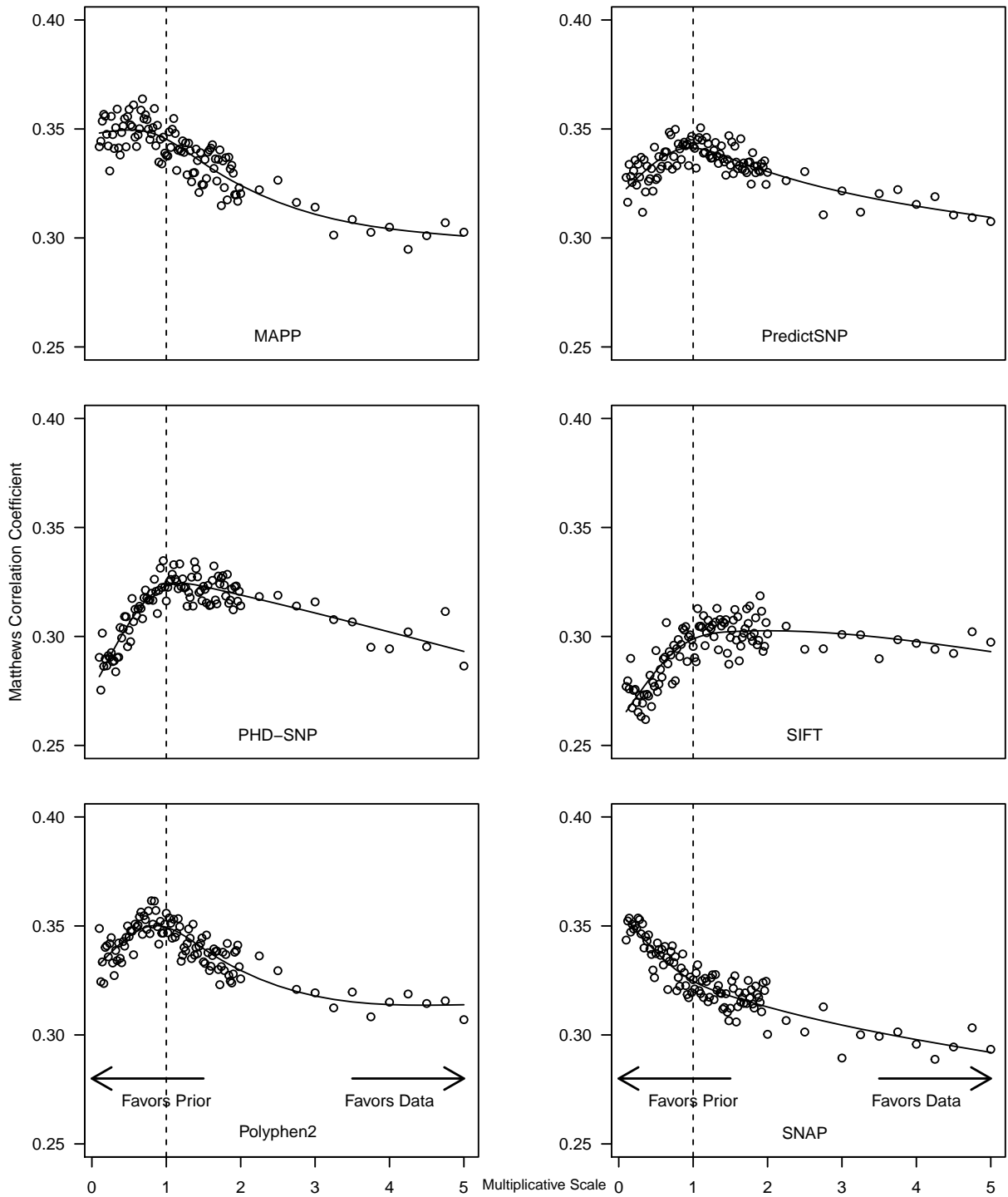

Figure 4: Cross-validated Matthews Correlation Coefficient given different degrees of regularization. Scale multipliers near zero are highly regularized whereas large multipliers represent very little regularization.

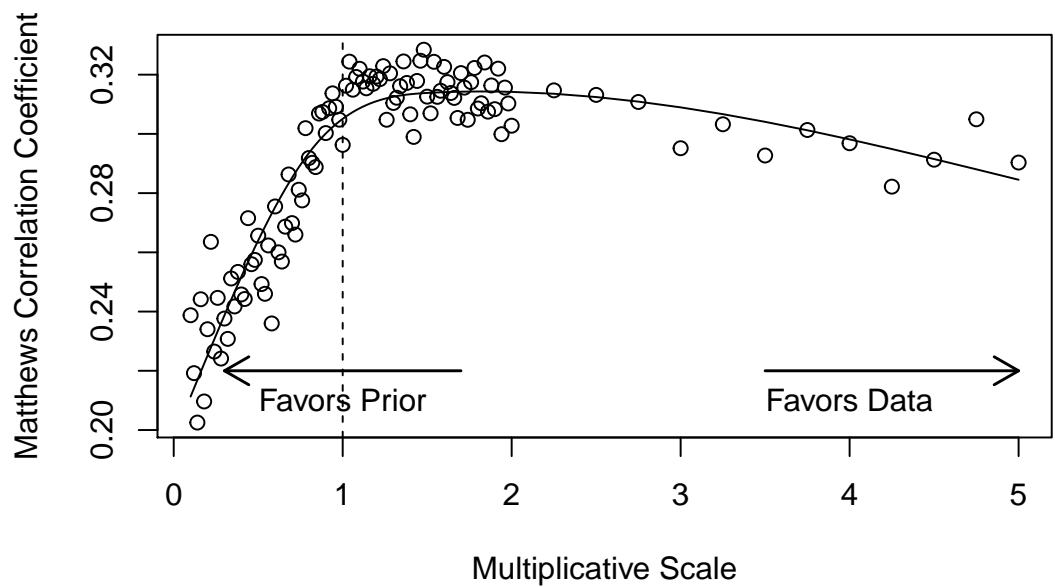

Figure 5: Result of the empirical Bayes regularization procedure when the input model is composed of random draws from a Unif(0,1). As expected the optimal model still performs reasonably with emphasis being placed on the data rather than the artificial prior.

# 5 NSQIP Example

## 5.1 Dataset

196 observations were taken from a clinical trial on the efficacy of statins for the prevention of perioperative kidney injury in patients undergoing cardiac surgery. Each patient had available data for age, gender, emergency surgery, history of diabetes, history of hypertension, history of congestive heart failure, history of chronic obstructive pulmonary disease, smoking status, baseline serum creatinine levels and CPT code of the surgery being performed. This information was used to obtain each patient's risk of any surgical complication according to the NSQIP online surgical risk prediction tool [13]. The outcome of interest for these patients is whether the patient experienced a serum creatinine increases greater than or equal to 0.3 mg/dL in the first two postoperative days.

## 5.2 Statistical Modeling

The same procedure was used in this example was used for the KCNQ1 data. As can be seen in figure 6, the model again benefited from the regularization provided by the empirical Bayes procedure, with the optimum performance again being in the vicinity of the baseline scale. In this example it is even more evident that the NSQIP model did not contain as much applicable information. There is a steep decline in performance as more weight is placed on the extracted prior.

Once again the model was compared to an optimized elastic net regression using the same cross-validation procedure. In this case the optimal parameter values were at ($\alpha = 1, \lambda = 0.033$), making the optimal elastic net equivalent to the optimal LASSO model. This model had a cross-validated MCC of 0.14 making it very comparable but slightly less than the model fit using the empirical Bayes procedure.

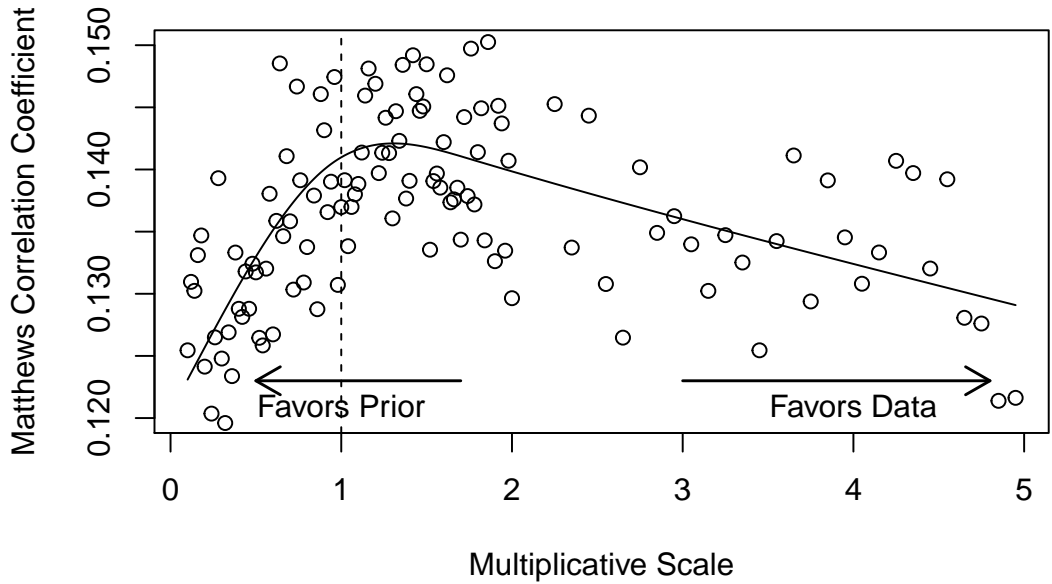

Figure 6: Matthews Correlation Coefficient as a function of the scaling factor applied to the baseline prior for the AKI dataset augmented with predictions from the NSQIP Surgical Risk score.

# 6 Discussion

Empirical Bayes regularization provides a compelling alternative to penalized regression as well as fully Bayesian models that require specification of a prior, often from a state of ignorance. By conducting shrinkage toward a plausible value for the $\beta$s instead of to zero, the empirical Bayes method puts a lower bound on the optimal predictive performance that is at least as high or higher than that achieved by penalized regression. Prior to fitting the penalized regression model the analyst can be confident that resulting optimized performance will be at least as good as the unpenalized model since the optimization procedure can select a penalty of zero and return the unpenalized model. Similarly the empirical Bayes approach can select an extremely diffuse prior and essentially invoke no regularization. However, the empirical Bayes approach can also select to return a fully regularized model and return the prior predictive estimates. Therefore, the lower limit of performance on the empirical Bayes procedure is higher than that given by regularization toward zero. If $\gamma$ is the chosen metric of performance then

$$\gamma_{unpen} \leq max(\gamma_{unpen}, \gamma_{prior})$$

Although this does not guarantee superior performance of the empirical Bayes method in any particular case, the worst possible result from the regularization procedure is necessarily better.

In all seven examples the optimal empirical Bayes procedure was at least as good as elastic net regularization. Because ridge and LASSO regression are nested within the elastic net model this can also be taken as evidence that the empirical Bayes method is capable of exceeding the performance of each of these regularization methods.

The "prior" extraction process via the parametric bootstrap provides a general framework for establishing both a centrality parameter and a baseline scale to be used in the empirical Bayes procedure, although the latter is less important due to the tuning process. Whether the scale of the estimated prior is treated as a tuning parameter or the problem is framed as modified $L^p$ regularization where $||\beta - \bar{\beta}||_p$ is penalized, the result is decreased variance with the applied bias being in the direction of a meaningful value. This approach displays an amalgamation of the best properties of the fully Bayes and penalized regression approaches.

An additional layer of complexity can be considered when performing this sort of regularization. In this manuscript it has been recommended that the results of the parametric bootstrap be used to establish a baseline scale for the prior distribution. This baseline scale turns finding the optimal predictive performance into a one dimensional problem, i.e. a single tuning parameter akin to ridge regression or LASSO. A multiplicative factor is used to scale each baseline prior by the same amount. It is likely that this procedure will not result in globally optimum predictive performance as regularizing the various coefficients to different degrees may improve performance. This approach was not pursued as it is unlikely that any minor improvement in performance would be worth a substantial increase in the number of dimensions over which the solution must be optimized. This setup would be more analogous to the non-negative garrote as opposed to $L^p$ type regularization.

# 7 Summary

Previously published predictive models should be leveraged even when it is not evident that it is directly applicable to the problem at hand. The empirical Bayes regularization will naturally place emphasis on whatever portions of the model result in the best predictive performance. Due to the generality of the empirical Bayes approach and to shrinkage occurring toward a plausible value rather than toward zero, the empirical Bayes approach provides an attractive alternative to other penalization schemes.